\documentclass[twocolumn,prl,showpacs,10pt,a4paper]{revtex4}%
\usepackage{epsfig,amssymb,amsfonts}
\usepackage{amsmath}
\usepackage{graphicx}
\begin{document}
\title{Universal and deterministic manipulation of the quantum
state of harmonic oscillators: a route to unitary gates for Fock State qubits}
\author{Marcelo Fran\c{c}a Santos}
\email{msantos@fisica.ufmg.br}
\affiliation{Dept. de F\'{\i}sica, Universidade
Federal de Minas Gerais, Belo Horizonte, 30161-970, MG, Brazil}
\begin{abstract}
We present a simple quantum circuit that allows for the universal and
deterministic manipulation of the quantum state of confined harmonic
oscillators. The scheme is based on the selective interactions of the referred
oscillator with an auxiliary three-level system and a classical external
driving source, and enables any unitary operations on Fock states, two-by-two.
One circuit is equivalent to a single qubit unitary logical gate on Fock states
qubits. Sequences of similar protocols allow for complete, deterministic and
state-independent manipulation of the harmonic oscillator quantum state.
\end{abstract}


\maketitle

In the last two decades, state-of-the art experiments on both cavity
QED~\cite{haroche} and trapped ions~\cite{Wineland} have been exploring the
quantized nature of different spatially confined harmonic oscillators. As
examples of important experimental demonstrations we can cite the production
and detection of Fock states~\cite{1foton,2fotons}, Schroedinger-cat like
states~\cite{cat,monroe}, as well as the complete measurement of non-classical
quasi-probability distributions in quantum phase
space~\cite{1fotonqed,wineland2}.

Recently, understanding and operating on those quantized harmonic oscillators
has also become an important issue for quantum information theory~\cite{Chuang}
both from a fundamental point of view as well as from practical
implementations. Quantized light and vibrational modes play essential roles in
many different proposals and experiments of quantum protocols, be it as
memory~\cite{haroche}, information buses~\cite{cirac} or even as computational
qubits~\cite{zeilinger,KLM}.

The possibility to investigate in a more controllable fashion different
quantized confined harmonic oscillators and to use them as qudits has prompted
a renewed theoretical and experimental effort towards the engineering and
manipulation of more complex quantum states in these systems~\cite{propostas}.
These proposals and experiments are based on coupling the quantized harmonic
oscillator to classical sources and low-dimensional systems such as few
selected electronic levels of neutral atoms or ions. All of them explore
important features of quantized harmonic oscillators. However, none of them
allows for a complete, state-independent, universal manipulation of quantum
states in these systems. In fact, even the apparently simple task of rotating a
Fock state qubit remained solvable only in the $\{|0\rangle,|1\rangle\}$
subspace.

In this paper, we combine those two interactions and the concept of
spin-echo~\cite{spinecho} to present a quantum circuit that implements any
state-independent, unitary transformation on arbitrary quantum states of
harmonic oscillators. Our proposal is based on the so-called selective
interactions, described in~\cite{mkr,amv} for Cavity QED and
in~\cite{kike,kike1} for trapped ions. In these interactions, an external
classical source is used to control the effective coupling between the harmonic
oscillator and auxiliary electronic levels of a neutral atom or an ion. The
scheme here presented is general and can be applied to any harmonic oscillator
in the presence of an external driving classical source and an auxiliary
three-level system.

In the first part of this manuscript we show how to combine two selective
interactions and an intermediate spin flip to build an universal deterministic
logical gate for a chosen Fock state qubit. In other words, we show how to
implement an arbitrary rotation in a chosen two-dimensional
$\{|m\rangle,|m-1\rangle\}$ energy subspace of the harmonic oscillator. This
three-step circuit (see Fig.1), which we will refer to as $\textrm{UG}_{m}$
will, then, constitute the fundamental block for arbitrarily manipulating any
state of the harmonic oscillator, i.e. sequences of $\textrm{UG}_{m}$ gates,
adjusted to rotate different pairs of Fock states (different "m's''), can be
used to manipulate more general states of the harmonic oscillator. Note that in
order to illustrate the scheme for practical applications, we analyze it in the
context of Cavity QED, but always keeping in mind that similar setups are
available for manipulating the vibrational modes of trapped ions.

The building block of our circuit involves two types of operations as shown in
Fig(1): a selective coupling $\widehat{H}(\theta)$ between two Fock states of
the harmonic oscillator (in our example, a cavity mode) and an auxiliary low
dimensional system (a three-level atom) intercalated by a local operation on
the auxiliary system. By selective, we mean a linear interaction
$\widehat{H}(\theta)$, between the atom and the light field that is resonant
for a chosen joint subspace and dispersive for all the remaining ones. For
example, if we denote the atomic states by $\{|g\rangle,|e\rangle\}$ and we
choose a particular Fock state $m$ for the cavity field, then
$\widehat{H}(\theta)$ is engineered so that only states
$\{|gm\rangle,|em-k\rangle\}$ perform Rabi oscillations. All the other doublets
evolve dispersively and all the remaining states just acquire
phases~\cite{mkr}. We show bellow how selectivity allows for the rotation of an
arbitrary harmonic oscillator qubit $|qb \rangle = \alpha |m-1 \rangle + \beta
|m \rangle$. The extension to more general two-dimensional subspaces
$\{|m\rangle,|m-k \rangle\}$ is immediate for the vibration of trapped ions.

First, let us briefly summarize the selective interaction in cavity QED. Much
like in~\cite{mkr} let us consider an off-resonance Raman Hamiltonian
described, in the interaction picture, by $\widehat{H}_{\mathrm{int}}=\hbar
g\widehat{\sigma }_{hg}\widehat{a}e^{-i\delta
t}+ \hbar \Omega _{L}\widehat{\sigma }_{he}e^{-i\delta t}+%
\mathrm{H.c.}$ (see Fig(2) for the levels scheme). The first term couples
dispersively, with coupling constant $g$, the lowest atomic energy level $|g
\rangle$ to the higher one $|h \rangle$ through the quantized mode described by
annihilation operator $\hat{a}$. The second term couples dispersively, with
coupling constant $\Omega_L$, the intermediate atomic energy level $|e \rangle$
with level $|h \rangle$ through an external driving source (for example, an
intense laser field). $\delta =\omega _{hg}-\omega _{0}=\omega _{he}-\omega
_{L}$ is the detuning between both atomic transitions and their respective
interacting fields frequencies.

When $\delta \gg |\Omega _{L}|,g$ (considering $g$ real), we can adiabatically
eliminate level $\left| h\right\rangle $ and approximate
$\widehat{H}_{\mathrm{int}}$ by the effective Hamiltonian ($\hbar
=1$)~\cite{mkr}:
\begin{equation*}
\widehat{H}_{\mathrm{eff}}(\theta)=\frac{g^{2}\widehat{a}^{\dagger
}\widehat{a}}{\delta }\widehat{\sigma }_{gg}+ \frac{g^2 m}{\delta
}\widehat{\sigma }_{ee}+ \lambda(e^{i\theta}\widehat{\sigma
}_{ge}\widehat{a}^{\dagger }+e^{-i\theta} \widehat{\sigma}_{eg}\widehat{a}),
\end{equation*}
where $\theta$ is the phase of $\Omega_L$, $\lambda = \frac{g|\Omega _{L}^{\ast
}|}{\delta }$ and we have already included an energy shift $\Delta_{m} =
\frac{g^2 m - |\Omega_L|^2}{\delta}$ to level $|e \rangle$ implementable
through the action of an external classical source, where $m$ is an integer
number. This Hamiltonian splits the joint atom-cavity mode Hilbert space into
two-dimensional subspaces spanned by the doublets $\{|g,n
\rangle,|e,n-1\rangle\}$. It describes a dispersive (or resonant) dynamics
whenever $\Delta(n) \gg \lambda$ (or $\Delta(n) \leq \lambda$), just like the
typical Jaynes-Cummings interaction~\cite{JC}. However, unlike the JC model,
now, the effective detuning $\Delta(n) = \frac{g^{2}(n - m)}{\delta}$ between
levels $|g,n \rangle$ and $|e,n-1 \rangle$ depends on the number $n$ of
excitations in the cavity field. Selectivity is achieved when $\Delta(n) \gg
\lambda$ for all $n \neq m$, which means that all the doublets evolve
dispersively except for a chosen one $\{|g,m\rangle,|e,m-1\rangle \}$ that
evolves resonantly ($\Delta(m)=0$). In our case, selectivity holds for $g \gg
|\Omega_L|$.

In the selective regime, second order Hamiltonian
$\widehat{H}_{\mathrm{eff}}(\theta)$ unfolds as $\widehat{H}_{0\rm{n}} +
\widehat{H}(\theta)$. The first part, given by
\begin{equation}
\widehat{H}_{0\rm{n}}= \sum_{n \neq m,m-1} \left[\frac{g^2 n }{\delta}|g\rangle
\langle g| + \frac{g^2 m }{\delta}|e\rangle \langle e|\right]\otimes |n \rangle
\langle n|, \label{dispersive}
\end{equation}
describes the dispersive dynamics for all the doublets that do not involve Fock
states $\{|m\rangle,|m-1\rangle\}$, whereas $\widehat{H}(\theta) =
\widehat{H}_{0\rm{m}}+ \widehat{H}_c(\theta)$ describes the interaction between
these Fock states and the atomic levels. $\widehat{H}_{0\rm{m}}$ contains the
self-energy correction terms for the joint states
$\{|g,m\rangle,|g,m-1\rangle,|e,m\rangle,|e,m-1\rangle \}$,
\begin{equation}
\widehat{H}_{0\rm{m}}=\frac{g^2 m }{\delta} \mathbb{I}_s - \frac{g^2 }{\delta}
|g,m-1\rangle \langle g,m-1|,
\end{equation}
where $\mathbb{I}_s$ is the identity in this subspace, and,
\begin{eqnarray}
\widehat{H}_c(\theta)=\lambda \sqrt{m}(e^{i\theta}|g,m\rangle \langle e,
m-1|+h.c.),
\end{eqnarray}
describes the selective coupling itself. Note that $\widehat{H}_c(\theta)$
allows for excitation exchanges between the atom and the harmonic oscillator to
happen only inside the chosen subspace $\{|g,m\rangle,|e,m-1 \rangle\}$. As a
consequence, if the harmonic oscillator is initially prepared in a
superposition of Fock states $|m\rangle$ and $|m-1\rangle$, no other Fock state
gets populated during the interaction with the atom. This property turns out to
be the only necessary condition for Fock states qubits deterministic
manipulation.

Let us consider an initial product state $|\Psi\rangle = |\psi_{at}\rangle |qb
\rangle$ between the atom and the cavity mode, where the atom is prepared in
the symmetric (or anti-symmetric) superposition of its internal electronic
states, $|\psi_{at}\rangle =\frac{ |g\rangle \pm |e\rangle}{\sqrt{2}}\equiv
|\pm \rangle$. This choice for the atomic state is justified later on when it
also becomes clear that both symmetric and anti-symmetric states are equally
good for this single qubit rotation protocol. Now, let us analyze in details
the following sequence of operations: first the atom interacts selectively with
the harmonic oscillator for a chosen time $\tau$. Then an external driving
field flips the atomic state (spin-echo technique), and finally atom and
harmonic oscillator interact selectively again for the same time $\tau$ but at
a slightly rotated angle ($\theta_0 = \frac{g^2}{\Delta}\tau$). The final joint
state is, then, given by:
\begin{equation}
|\Psi_f\rangle = e^{-i\widehat{H}(\theta_0)\tau}\sigma_x
e^{-i\widehat{H}(0)\tau}|\psi_{at}\rangle |qb \rangle. \label{estadof},
\end{equation}
where $\widehat{H}(\theta)= \widehat{H}_{0m}+\widehat{H}_c(\theta)$ and
$\sigma_x = |g\rangle\langle e| + |e\rangle\langle g|$.

In order to rewrite this time evolution in a clearer version, first note that
$\sigma_x e^{-i\widehat{H}(0)\tau}\sigma_x = e^{-i\widehat{H}_x(0)\tau}$, where
\begin{eqnarray}
\widehat{H}_x(0) = \frac{g^2 m }{\Delta} I_s - \frac{g^2
}{\Delta} |e,m-1\rangle \langle e,m-1| \nonumber\\
+ \lambda \sqrt{m}(|e,m\rangle \langle g, m-1|+h.c.).
\end{eqnarray}
Also note that
$[\widehat{H}_{0m},\widehat{H}_c]=[\widehat{H}_c,\widehat{H}_{xc}]=0$ (where
$\widehat{H}_{xc} = \sigma_x \widehat{H}_{c} \sigma_x$) , and that
$e^{-i\widehat{H}_{0m}\tau}e^{-i(\widehat{H}_c(\theta_0)+
\widehat{H}_{xc}(0))\tau}e^{i\widehat{H}_{0m}\tau}=
e^{-i(\widehat{H}_c(\theta_0)+ \widehat{H}_{xc}(\theta_0))\tau}$. Using these
properties of the effective Hamiltonian and the Baker-Hausdorff theorem, we can
rearrange this product of exponentials as
\begin{equation}
|\Psi_f \rangle = e^{-i(\widehat{H}_c(\theta_0)+
\widehat{H}_{xc}(\theta_0))\tau} e^{-i(\widehat{H}_{x0m}+
\widehat{H}_{0m})\tau} |\psi_{at} \rangle |qb \rangle
\end{equation}
After simple algebra it is straight forward to show that
$(\widehat{H}_{x0m}+\widehat{H}_{0m})\tau=2 \eta \mathbb{I}_s + \theta_0
\mathbb{I}_{at}\otimes |m-1\rangle \langle m-1|$, where $\mathbb{I}_{at}$ is
the identity in the atomic subspace and $\eta = \frac{g^2 m }{\Delta}\tau$.
This part represents the addition of a global phase $\eta$ to state
$|\Psi\rangle$ and the addition of a small phase $\theta_0$ to the qubit state
$|m-1 \rangle$. It takes state $|\Psi\rangle$ into state $|\Psi'\rangle = e^{-2
i\eta}|\psi_{at}\rangle |qb' \rangle$, where $|qb' \rangle = \alpha
e^{-i\theta_0}|m-1 \rangle + \beta |m \rangle$. Similar calculations lead to
the relation
\begin{eqnarray}
\left[\widehat{H}_c(\theta_0)+\widehat{H}_{xc}(\theta_0)\right]\tau= \left
\{|+\rangle \langle +|- |-\rangle \langle -|\right \} \nonumber \\
\times \phi(e^{i\theta_0}|m\rangle \langle m-1|+h.c.),
\end{eqnarray}
where $\phi = \lambda \tau \sqrt{m}$. This part represents a direct linear
coupling between qubit states $|m \rangle$ and $e^{-i\theta_0}|m-1 \rangle$
provided the atom is initially prepared in an eigenstate of the described
$\sigma_x$ operator. It takes state $|\Psi'\rangle$ into the final state
$|\Psi_f\rangle = |\phi_{at}'\rangle |qb_f \rangle$, where
\begin{equation*}
|qb_f \rangle = (\alpha \cos{\phi} + i\beta \sin{\phi})e^{-i\theta} |m-1
\rangle + (\beta \cos{\phi} + i\alpha \sin{\phi}) |m \rangle.
\end{equation*}
As it is shown above, if the atom is initially prepared in an eigenstate of
$\sigma_x$, then the proposed three step process implements a unitary rotation
in the Fock states basis $\{|m\rangle,|m-1\rangle\}$ of the harmonic
oscillator, without entangling it with the atomic state. Note that after the
first interaction atom and harmonic oscillator may get highly entangled. Being
a local operation, the spin-flip does not change this degree of entanglement,
but the final two-qubit operation implements the final rotation disentangling
both systems.

Also note that if levels $|g \rangle$ and $|e \rangle$ have approximately the
same energy, for example, if they form a hyperfine structure of the same
electronic level, then the intermediate step of the spin flip can be replaced
by an energy shift in one (or both) of them inverting their roles in the
interaction, i. e. producing an anti-Jaynes-Cummings coupling between these
levels and the quantized mode, similar to the one proposed in ~\cite{mkr}.

\begin{figure}
\centerline{\psfig{figure=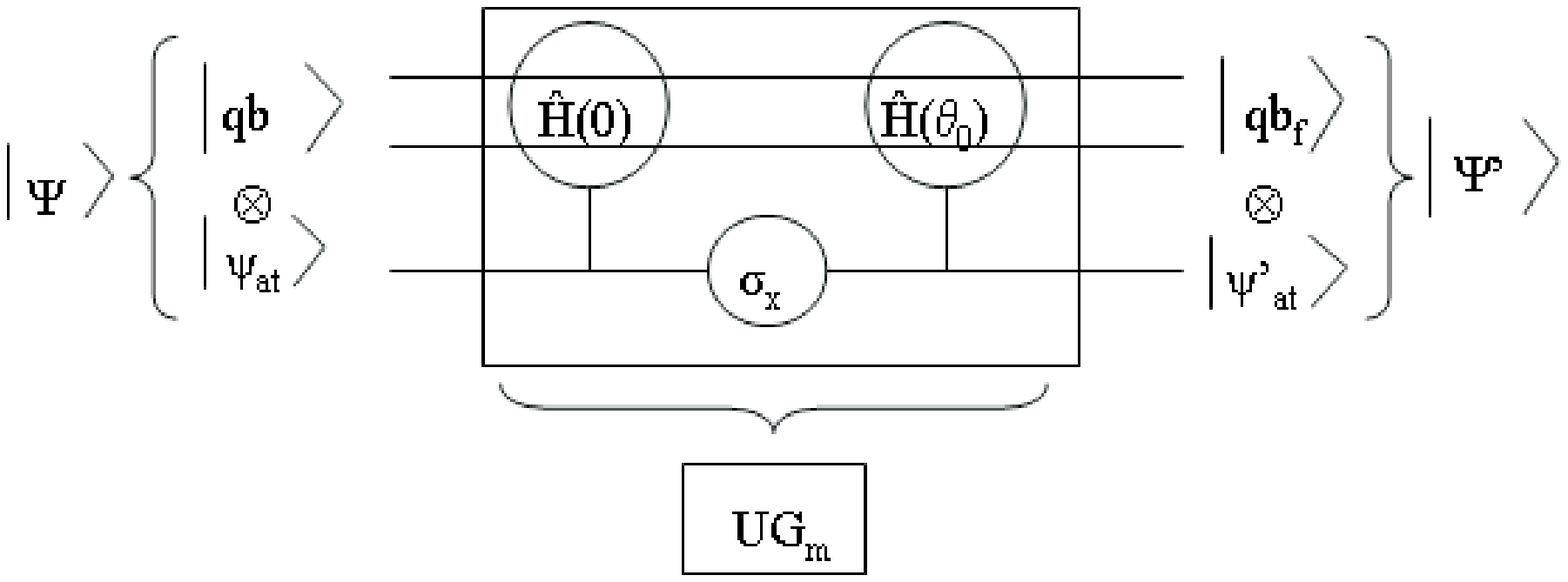, height=4.1 cm, width=8. cm}}
\caption{Primary circuit: three steps quantum circuit that rotates Fock states
$\left \{|m\rangle,|m+1\rangle \right\}$ into $\left \{\cos \theta |m\rangle +
e^{i\phi}\sin\theta |m+1\rangle, -\sin\theta
|m\rangle+e^{-i\phi}\cos\theta|m+1\rangle \right\}$} \label{fig1}
\end{figure}

\begin{figure}
\centerline{\psfig{figure=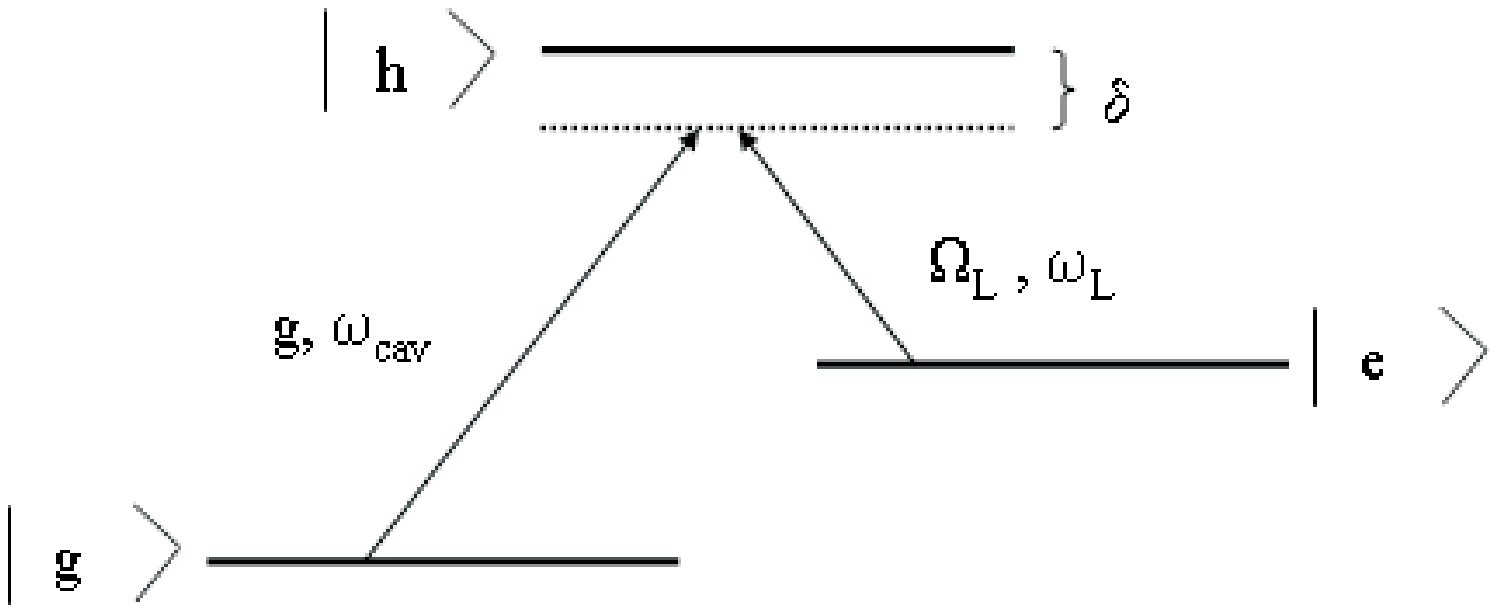, height=4.1 cm, width=10. cm}}
\caption{Atomic level scheme: lower-level $|g\rangle$ is dispersively coupled
to higher-level $|h\rangle$, through the cavity mode $\omega_{cav}$, with
coupling constant $g$, while intermediate level $|e\rangle$ is dispersively
coupled to level $|h\rangle$ through the external source field $\omega_L$ with
coupling constant $\Omega_L$.} \label{fig2}
\end{figure}

The same selectivity is also available in the trapped ions setup. For example,
Ref.~\cite{kike} shows how to produce the same effective Hamiltonian in this
context. In fact, due to the nature of the coupling between the internal levels
of a trapped ion and its vibrational motion, in the resolved sideband regime,
it is possible to engineer selectivity for the so-called
multi-quantum-Jaynes-Cummings model~\cite{kike1}, described by the coupling
term:
\begin{equation*}
\widehat{H}_{\textrm{int}k}(\theta)=\lambda_k(\widehat{\sigma
}_{ge}\widehat{a}^{\dagger k }+e^{-i\theta} \widehat{\sigma}_{eg}\widehat{a
}^k),
\end{equation*}
In this case, the selected doublet is $\{|g,m\rangle,|e,m-k\rangle\}$.
%reads $\Delta(n) \gg \lambda_k
%\sqrt{\frac{n!}{(n-k)!}}$ and the resonant doublet is

Now that we showed a quantum gate that rotates arbitrary Fock states qubits, it
is trivial to extend the idea to higher dimensional qudits in order to engineer
and manipulate any state of quantized harmonic oscillators. Note that the basic
circuit $UG_m$ is a combination of two and one qubit gates. In fact, what we
present next is a practical application of the known fact that any quantum
computation can be executed with those two elements~\cite{divincenzo}.

Given any initial state $|\Phi\rangle = |\psi_{at}\rangle \sum_{n} c_n
|n\rangle$, one can produce any other state by applying rotations involving
different subspaces $\{|m\rangle,|m-k\rangle\}$, each one of them selected by
its respective shift $\Delta_m$ to atomic level $|e\rangle$. Fig.(3) shows the
quantum circuit for this qudit manipulation using only one atomic qubit,
operating on different pairs of Fock states, one pair at a time. Note that the
whole operation can be done with only one auxiliary system, due to the fact
that after each one qubit gate there is no entanglement between the harmonic
oscillator and the auxiliary system. For example, beginning with the harmonic
oscillator in its ground state $|0\rangle$, one can use a sequence of $n$
gates, to prepare the superposition $\alpha |0\rangle + \beta |n\rangle$. First
we shift level $|e\rangle$ by $\Delta_1 = \frac{g^2}{\Delta} -
\frac{|\Omega_L|^2}{\Delta}$ corresponding to the primary circuit
$\textrm{UG}_1$ which operates on the subspace $\{|0\rangle,|1\rangle\}$ of the
harmonic oscillator. This first gate is used to prepare the quantum
superposition $\alpha|0\rangle+\beta|1\rangle$. Then, we select the energy
shift $\Delta_2$ to coherently transfer the population of Fock state
$|1\rangle$ to Fock state $|2\rangle$, with no changes to the population of the
ground state, since $\textrm{UG}_2$ operates only on the subspace
$\{|1\rangle,|2\rangle\}$. This second operation prepares state
$\alpha|0\rangle+\beta|2\rangle$. It is clear that the sequence of operations
$\textrm{UG}_n(\tau_n)...\textrm{UG}_2(\tau_2)\textrm{UG}_1(\tau_1)|0\rangle$,
where $\alpha = \cos{\frac{g |\Omega_L^*| \tau_1}{\Delta}}$  ($\beta =
\sin{\frac{g |\Omega_L^*| \tau}{\Delta}}$) and $\frac{\sqrt{j}g |\Omega_L^*|
\tau_{j\neq 1}}{\Delta}=\frac{\pi}{2}$ prepares the quantum superposition
$\alpha|0\rangle + \beta|n\rangle$.

\begin{figure}
\centerline{\psfig{figure=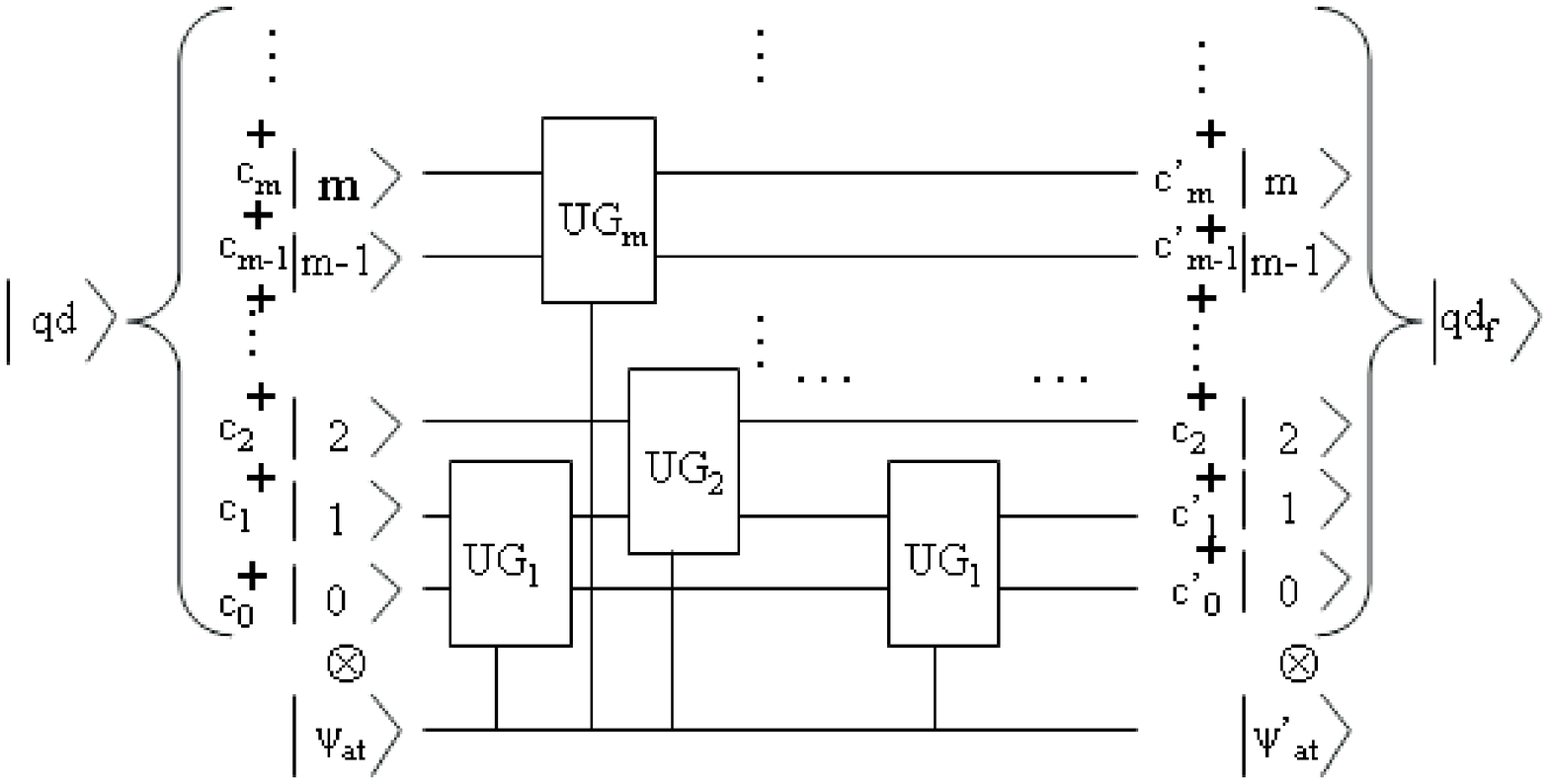, height=6.cm, width=10. cm}}
\caption{Quantum circuit that implements deterministic transformations in
arbitrary states of the harmonic oscillator, taking state $|qd\rangle = \sum_n
c_n |n\rangle$ into state $|qd_f\rangle = \sum_n c_n' |n\rangle$ using just one
auxiliary system. In this case, the short circuits $UG_m$ must be sequential in
time} \label{fig4}
\end{figure}

In this particular example, the gates must be sequential in time, i.e. first we
operate $\textrm{UG}_1$, then $\textrm{UG}_2$ and so forth. However, depending
on the desired transformation, more than one auxiliary qubit can be used in
order to accelerate the process, since each one of them can be adjusted (by
different shifts) to rotate a particular subspace of the harmonic oscillator
qudit, as shown in Fig (4). In this case, if $N$ is the higher Fock state to be
manipulated, then $N/2$ auxiliary systems can be used to fasten the whole
operation. This second approach is particularly interesting for the
manipulation of vibrations of trapped ions, given that all the ions in the same
trap are coupled to their different collective vibrational modes.

\begin{figure}
\centerline{\psfig{figure=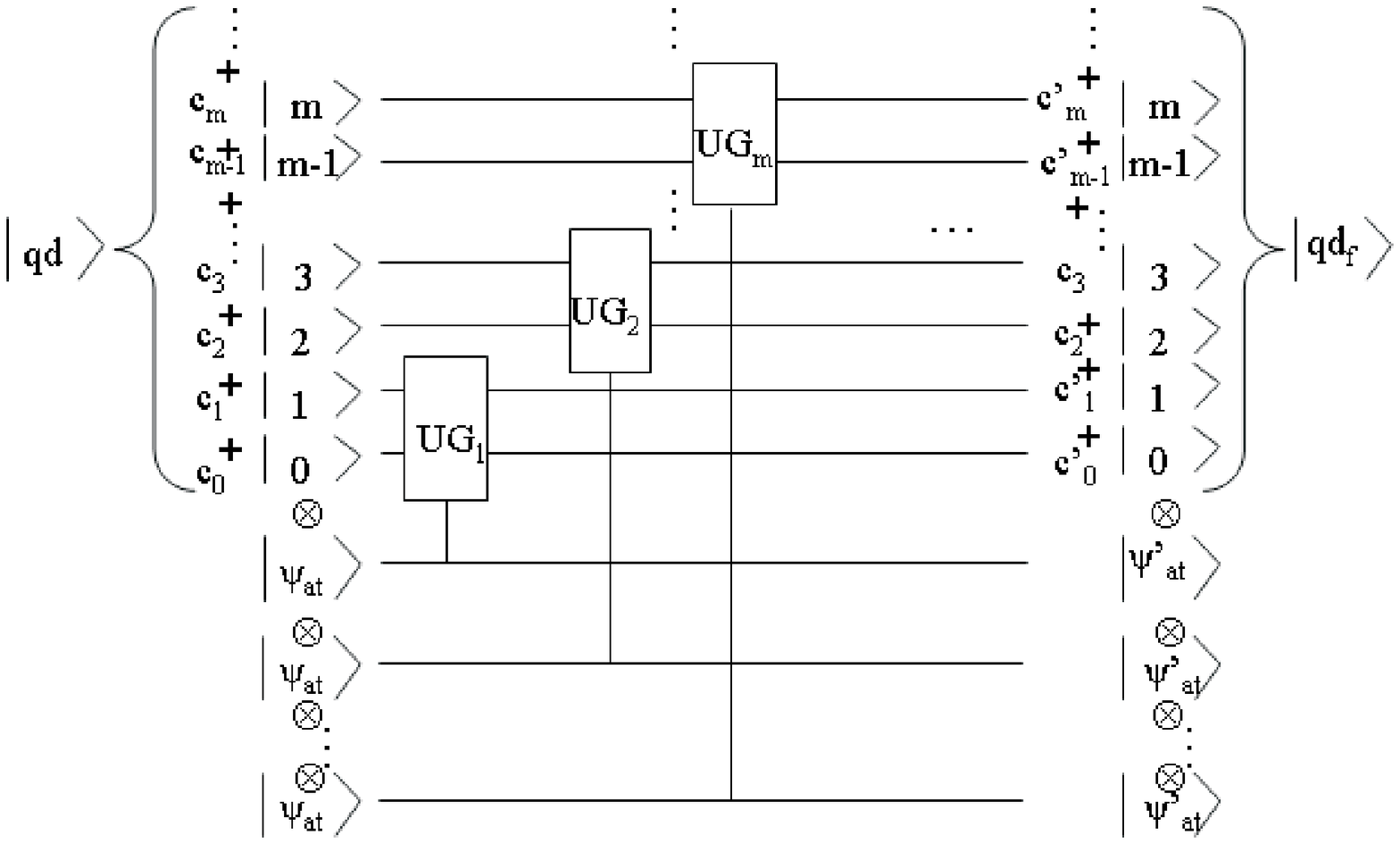, height=6.cm, width=10. cm}}
\caption{Quantum circuit that implements deterministic transformations in
arbitrary states of the harmonic oscillator, taking state $|qd\rangle =
\sum_n^N c_n |n\rangle$ into state $|qd_f\rangle = \sum_n^N c_n' |n\rangle$
using $N/2$ auxiliary systems. In this case, $N/2$ short circuits $UG_m$ can be
processed in parallel, since each auxiliary system can be used to manipulate a
pair of consecutive Fock states at the same time.} \label{fig5}
\end{figure}

In this manuscript, we present a simple quantum circuit, based on a selective
interaction, that implements arbitrary unitary transformations on Fock states
qubits. We also show that sequences of this circuit can be used to operate any
unitary transformation on the quantum state of harmonic oscillators.

\begin{acknowledgments}
The author acknowledges the support of CNPq and thanks M.O. Terra Cunha, C.
Villas-Boas and L. Sanz for useful discussions.
\end{acknowledgments}


\begin{thebibliography}{99}                                                                                               %

\bibitem{haroche} J. M. Raymond, M. Brune, S. Haroche, Rev. Mod. Phys.
\textbf{73}, 565 (2001).
%revisao de foton como qubit em QED

\bibitem {Wineland} D. Liebfried, R. Blatt, C. Monroe and D. Wineland, Rev. Mod. Phys., \textbf{75}, 281 (2003)
%resumo de ions armadilhados

\bibitem{1foton} G. Nogues, A. Rauschenbeutel, S. Osnaghi, M. Brune, J. M.
Raimond and S. Haroche, Nature \bf 400\rm, 239 (1999)
%deteccao de 1 foton na cavidade QND

\bibitem{2fotons} B.T.H. Varcoe, S. Brattke, M. Weidinger, and H. Walther,
Nature \bf 403\rm, 743 (2000)
%2 photons Fock state

\bibitem{cat} E. Schr\"{o}dinger, Naturwissenschaften {\bf 23},
807(1935).
%Schroedinger cat

\bibitem{monroe} C. Monroe, D.M. Meekhof, B.E. King, D.J. Wineland, Science {\bf 272} (5265): 1131-1136
(1996).

\bibitem{1fotonqed} P. Bertet, A. Auffeves, P. Maioli, S. Osnaghi, T. Meunier, M. Brune,
J. M. Raimond and S. Haroche, Phys. Rev. Lett. {\bf 89}, 200402 (2002).
%Wigner function one photon Fock state

\bibitem{wineland2} D. Leibfried, D.M. Meekhof, B.E. King, C. Monroe, W.M. Itano, D.J.
Wineland, Phys. Rev. Lett. {\bf 77}, 4281-4285 (1996).
%Wigner function 1 phonon Fock state

\bibitem{Chuang}  For a review on quantum protocols see M.A. Nielsen and I.L.
Chuang, {\it Quantum Computation and Quantum Information}. (Cambridge
University Press, Cambridge, 2000); {\em The physics of quantum information :
quantum cryptography, quantum teleportation, quantum computation} Springer, New
York, 2000, Dirk Bouwmeester, Artur K. Ekert, Anton Zeilinger (eds.).

\bibitem{cirac} J. I. Cirac, et al. Phys. Rev. Lett. \textbf{78}, 3221 (1997).
%computacao usando fotons como carregadores da informacao.

\bibitem{KLM}
E. Knill, R. Laflamme \& G.J. Milburn, {\it Nature} {\bf 409}, 46-52 (2001).

\bibitem{zeilinger} S. Gasparoni, J.W. Pan, P. Walther, T. Rudolph, A. Zeilinger,
Phys. Rev. Lett. {\bf 93} 020504 (2004).

\bibitem{propostas} see, for example: C. K. Law and J. H. Eberly Phys. Rev. Lett. {\bf 76}, 1055-1058 (1996),
L. Davidovich, M. Brune, J.M. Raimond, and S. Haroche, Phys. Rev. A \bf 53\rm,
1295 (1996); S.–C. Gou, J. Steinbach, and P. L. Knight Phys. Rev. A {\bf 54},
4315-4319 (1996); M. Dakna et. al, Eur. Phys. J. D, \textbf{3}, 295-308 (1998),
C. J. Villas-Bôas et al., Phys. Rev. A 68, 061801(R) (2003);
%schroed. cat in QED

\bibitem{spinecho} E. L. Hahn, Phys. Rev. {\bf 80}, 580–594 (1950)

\bibitem{mkr} M. Fran\c{c}a Santos, E. Solano, R. L. de Matos Filho, Phys. Rev.
Lett. \textbf{87}, 093601-1 (2001);

\bibitem{amv} A. Carollo, M.F. Santos, V. Vedral, Phys. Rev. A {\bf 67}, 063804 (2003).

\bibitem{kike} T. Pellizzari and H. Ritsch, Phys. Rev. Lett. \bf 72\rm, 3973 (1994); E. Solano,
P. Milman, R. L. de Matos Filho, and N. Zagury, Phys. Rev. A {\bf 62},
021401(R) (2000); H. Nha, Y.-T. Chough, and K. An, Phys. Rev. A, \bf 63\rm,
010301~R (2001);

\bibitem{kike1} E. Solano, quant-ph/0310007.
%selective interactions

\bibitem{JC} E.T. Jaynes and F.W. Cummings, \it Proc. IEEE \rm \bf 51\rm,
89 (1963).

\bibitem{divincenzo} D. P. DiVincenzo, Phys. Rev. A {\bf 51}, 1015-1022 (1995)
\end{thebibliography}
\end{document}